# Randomness and Earth's climate variability

Michael E. Levinshtein
Ioffe Institute, 26 Politekhnicheskaya, 194021 St. Petersburg, Russia
E-mail: melev@nimis.ioffe.ru

Valentin A. Dergachev
Ioffe Institute, 26 Politekhnicheskaya, 194021 St. Petersburg, Russia

Alexander P. Dmitriev
Ioffe Institute, 26 Politekhnicheskaya, 194021 St. Petersburg, Russia

Pavel M. Shmakov
Ioffe Institute, 26 Politekhnicheskaya, 194021 St. Petersburg, Russia

**Abstract.**

Paleo-Sciences including palaeoclimatology and palaeoecology have accumulated numerous records related to climatic changes. The researchers have usually tried to identify periodic and quasi-periodic processes in these paleoscientific records. In this paper, we show that this analysis is incomplete. As follows from our results, random processes, namely *processes with a single-time-constant* $\tau_0$ (noise with a Lorentzian noise spectrum), play a very important and, perhaps, a decisive role in numerous natural phenomena. For several of very important natural phenomena the characteristic time constants $\tau_0$ are very similar and equal to $(5-8) \times 10^3$ years. However, this value of $\tau_0$ is not universal. For example, the spectral density fluctuations of the atmospheric radiocarbon $\delta^{14}C$ are characterized by a Lorentzian with $\tau_0 \approx$ 300 years. The frequency dependence of spectral density fluctuations $S^{\delta^{18}O}$ for benthic $\delta^{18}O$ records contains two Lorentzians with $\tau_0 \approx 8000$ years and $\tau_0 > 10^5$ years.

# 1. Introduction.

The problem of studying and predicting the natural phenomena evolution of our planet is vital for all of us and for the Humanity as a whole. At present, there are steadily increasing changes of the natural conditions in various geospheres on the Earth's surface. These changes are caused by fluctuations of natural processes under the influence of heliocosmic and tectonic factors and possibly by the increasing human activity

It is widely believed that the rise in the average temperature on the Earth's surface in the past few decades is caused by the increasing human activity. Based on this assumption, trillions of dollars are spent to fight the global warming.

At the same time, the opponents of this concept argue that the periods of relative cooling and warming repeatedly alternated in the Earth's history without any human intervention. Suffice it to mention that only 20 000 years ago the whole territory of North Europe was occupied by a glacier with a thickness reaching 2.5 - 3 km. It is known for sure that approximately 9 – 10 thousand years ago this monstrous ice armour melted without any impact of human activity.

Apparently, the scientific prediction of climate changes on our planet is, putting it mildly, very important. The only scientific way to try to predict the future trends of these changes is by making a correct analysis of the climatic changes in the past. *The Paleo-Sciences* including palaeogeography, palaeoclimatology, and palaeoecology have accumulated over the years of studies tens of thousands of very different records related to climatic changes. Such important parameters as solar insolation, temperature measured in different areas in the scale of hundreds to millions of years, content of the oxygen isotope ($\delta^{18}O$), atmospheric radiocarbon ($\Delta^{14}C$), deuterium ($^2H$), etc., have been recorded and iteratively analyzed. Nevertheless, the relationships underlying the natural phenomena remain unexplored in many aspects. In particular, the researchers have usually tried to identify periodic and quasi-periodic processes in these paleoscientific records. In this paper, we show that this analysis is incomplete.

In this paper, we show that *random processes*, namely *single-time-constant random processes* (noise with a Lorentzian noise spectrum) play a very important and, perhaps, a decisive role in numerous natural phenomena.

## 2. Method and calculations

We consider a paleoenvironmental record of some quantity $\varphi$ as a random function ("noise") $\varphi(t)$. The spectral density of fluctuations $S^\varphi(f)$ is calculated as follows. Another random function, $\varphi'(t)$ is introduced defined as $\varphi'(t) = \varphi(t) - \overline{\varphi}$, where $\overline{\varphi}$ is the average value of $\varphi(t)$ over a long time interval $T$:

$$\overline{\varphi} = \frac{1}{T} \int_{-T/2}^{T/2} \varphi(t) dt \tag{1}$$

The average value of $\varphi'(t)$ is zero, so it describes the deviation of $\varphi(t)$ from the mean.

The spectral noise density of fluctuations $S^\varphi(f)$ can be expressed through the Fourier transform $\tilde{\varphi}(f)$ of the $\varphi'(t)$ function:

$$S^\varphi(f) = \frac{1}{T} |\tilde{\varphi}(f)|^2, \quad \tilde{\varphi}(f) = \int_{-T/2}^{T/2} \varphi'(t) \exp(i 2\pi f t) dt, \quad T \to \infty. \tag{2}$$

The function $S^\varphi(f)$ was calculated numerically.

## 3. Results and discussion

Paleoenvironmental records have, at first sight, the form of noise (so called "grass"). As an example, Figure 1 shows the time dependence of the virtual axial dipole moment (VADM) for last 800000 years (800 Ka) [1].

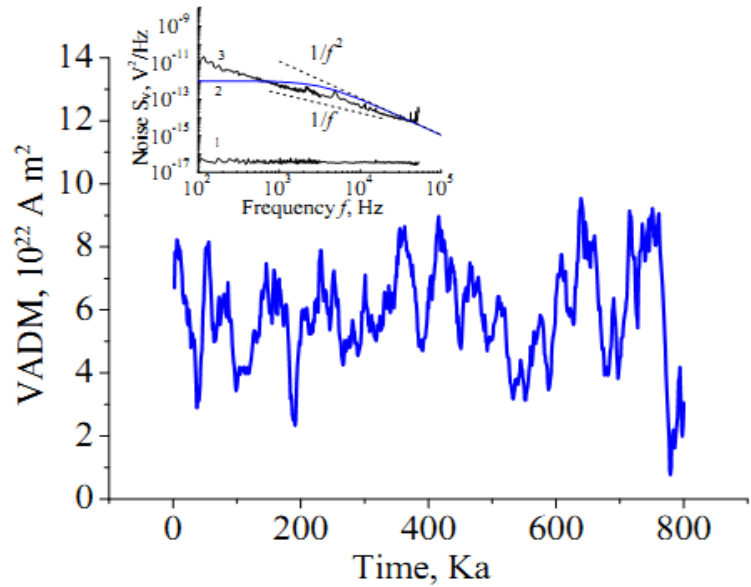

**Figure 1.** The time dependence of the virtual axial dipole moment (VADM) for 800 Ka [1]. Inset illustrates the most commonly occurring types of the frequency spectral noise density dependences: 1- white noise; 2 – noise with a *single time constant* (Lorentzian spectrum); 3 – the 1/f noise. The dependences are plotted for the case when the voltage fluctuations $\delta V$ are studied. Note logarithmic scale along both axes and the dimension of spectral noise density $S_v$ : [V$^2$/Hz]

Of course, periodic and quasi-periodic processes can be (and should be) identified in the dependence shown in Fig. 1. At the same time, this dependence can be regarded as "noise". In this case, the frequency dependence of the spectral density $S(f)$ of this "noise" should be calculated (see, for example, [2-4]). This function shows, roughly speaking, the frequencies at which the fluctuations are large and those at which they are small.

Studies of noise processes in various systems, materials, and devices reveal a variety of $S(f)$ dependences. However, in most cases the $S(f)$ dependence is represented by one of the three specific dependences, or a combination of these. The specific dependences are the so-called "white noise", "Lorentzian", and "1/f or flicker noise" (Inset in Fig. 1). When the white noise predominates (Inset in Fig. 1, curve 1) the spectral noise density is frequency-independent. The well-known examples of such dependence are the equilibrium thermal (Johnson-Nyquist) noise and the shot noise.

If the $S(f)$ dependence is described by a single Lorentzian (Inset in Fig. 1, curve 2), the frequency dependence of $S$ has the form:

$$S \propto \frac{\tau_0}{1+(2\pi f \tau_0)^2} \tag{3}$$

The well-known example of such dependence is generation-recombination noise in semiconductors and semiconductor devices [3,4]. In this case, only a random process with *one time constant $\tau_0 = 1/2\pi f_0$* contributes to the noise. As seen from (1), $S$ is frequency-independent at low frequencies, when $f \ll f_0$, and decreases as $1/f^2$ ($S \sim 1/f^2$) at $f \gg f_0$. Such a dependence $S(f)$ is quite often observed in semiconductors and metals when a single type of defects with high concentration is introduced into a material, for example, by doping or irradiation. It is noteworthy that the characteristic time constant $\tau_0 = 1/2\pi \times f_0$ is easily found from the $S(f)$ dependence: $f_0$ is the frequency at which the noise decreases relative to its low-frequency by a factor of 2 (3 dB)

The 1/f or (flicker) noise is observed in hundreds of objects: physical, chemical, biological, and even in results of human activities (see, for example, [5]). When the 1/f noise predominates, the spectral noise density decreases with increasing frequency as $S \sim 1/f^{\gamma}$, where $\gamma$ is close to unity. In most cases the nature of this noise in unknown, however, in the few cases where it is known, the noise is a superposition of closely spaced Lorentzians, i.e., superposition of multiple "elementary acts" with a very wide distribution of characteristic time constants $\tau_0$.

We calculated the frequency dependences of the spectral fluctuation density for several important paleoclimate records considering the time dependences of such parameters as VADM, temperature, the content of deuterium ($^2$H), oxygen isotope ($\delta^{18}$O), and atmospheric radiocarbon ($\Delta^{14}$C) as random processes ("noise")

Figure 2 presents the frequency dependence of the spectral density fluctuations $S^{DM}(f)$ of the virtual axial dipole moment (VADM) for 800 Ka (Fig. 1).

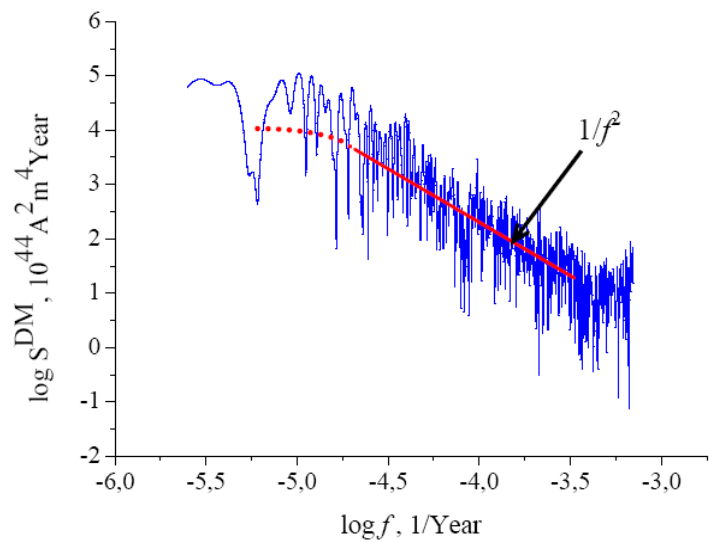

**Figure 2.** The frequency dependence of spectral density fluctuations, $S^{DM}$ for the time dependence virtual axial dipole moment (VADM) presented in Fig. 1. The solid and dotted lines are guidelines to the eye.

It can be seen in Fig. 2 that the overall run of $S^{DM}$ curve in the frequency range $10^{-5} \leq f \leq 10^{-3.5}$ 1/year is quite well described by a single Lorentzian with a characteristic time constant $\tau_0 \sim 10^{4.7}/2\pi \approx 8000$ years. According to the aforesaid, this means that, along with the possible periodic and quasi-periodic processes in the time dependence of VADM, an important (and possible dominant) influence is exerted *by a random process* with a characteristic time constant $\tau_0 \sim 8000$ years.

Figure 3a presents the frequency dependence of spectral density fluctuations, $S^{2H}$ for the deuterium content (Vostok Ice Core Deuterium Data for 420,000 years, [6]). Figure 3b shows a similar dependence for the deuterium data for 740,000 years from EPICA Dome C Ice Cores [7].

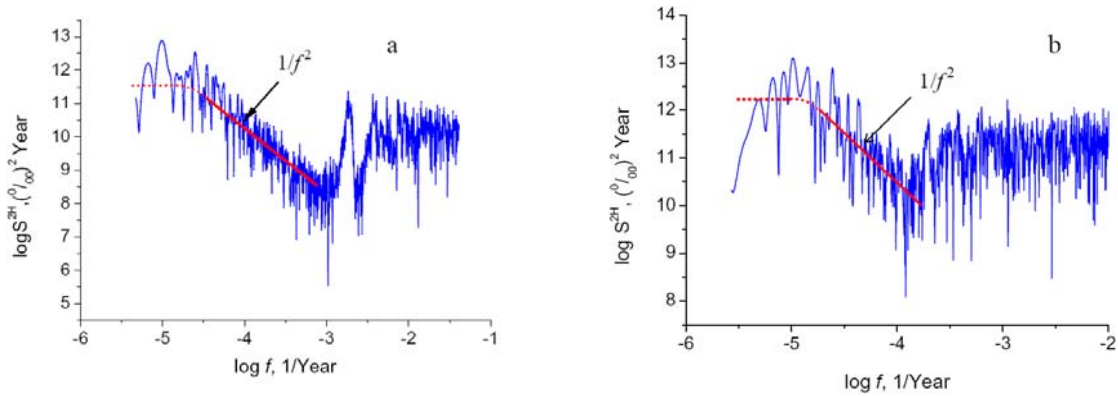

**Figure 3**. The frequency dependences of spectral density fluctuations, $S^{2H}$ for the deuterium content. a - Vostok Ice Core Deuterium Data for 420,000 years [6]; b - EPICA Dome C Ice Cores data for 740,000 years [7]. The solid and dotted lines are guidelines for the eye.

It is seen from Figure 3 that the overall behavior of $S^{2H}$ dependences in the frequency range $f \geq 10^{-5}$ 1/years is well described by a single Lorentzian with an characteristic time constant $\tau_0 \sim \frac{1}{2\pi} (10^{4.5}\text{-}10^{4.7}) \approx (5 \div 8) \times 10^3$ years. The fact that the values of $\tau_0$ are approximately the same for the data shown in Figs. 3a and 3b, is not surprising. Much more surprising is the apparent similarity between these dependences and $S^{DM}(f)$ dependence (Fig. 2).

It is interesting to compare the data presented in Figs. 2 and 3 with $S^{\delta 18O}(f)$ dependence for the content of oxygen $^{18}O$ isotope [8] (Figure 4)

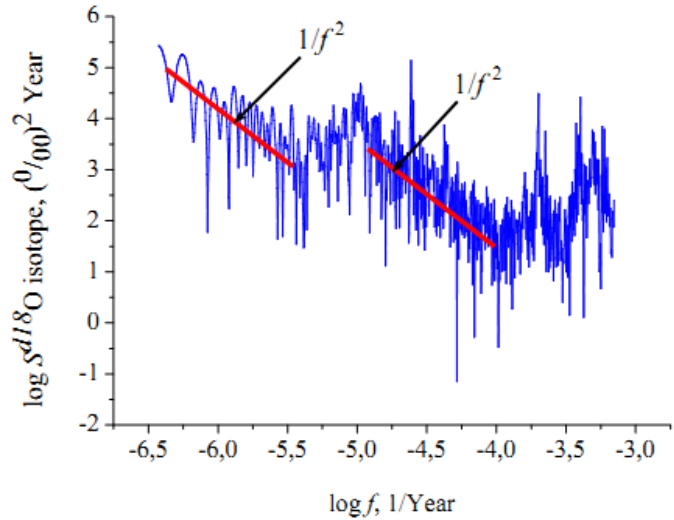

**Figure 4.** The frequency dependence of spectral density fluctuations $S^{\delta 18O}$ for globally distributed benthic $\delta^{18}O$ records [8]. The solid lines are guidelines for the eye.

The data presented in [8] span 5.3 Myr and are an average of 57 globally distributed benthic $\delta^{18}O$ records, which measure the global ice volume and the deep ocean temperature, collected from the scientific literature. It can be seen from Fig. 4 that there are two characteristic parts in the $S^{\delta 18O}(f)$ dependence, in which the spectral density $S^{\delta 18O}$ follows the law $S^{\delta 18O} \sim 1/f^2$. For the part in the frequency range $10^{-4.7} \leq f \leq 10^{-4.0}$ 1/year, the value of $\tau_0$ is close to the corresponding values of $\tau_0$, found for the plots shown in Figures 2 and 3. For the "low-frequency" part $10^{-6.3} \leq f \leq 10^{-5.3}$ 1/year, the value of $\tau_0$ cannot be found from the data presented. It is obvious, however, that the characteristic time $\tau_0$ of *this* random process, exceeds $10^5$ years.

A single Lorentzian with $\tau_0 \sim 10^{4.7}/2\pi$ years can be also clearly seen in Figure 5 which presents the frequency dependence of the spectral density of the temperature fluctuations $S^T(f)$ [9]. The dependence $S^T \sim 1/f^2$ is clearly observed in the frequency range $10^{-4.7} \leq f \leq 10^{-3.0}$ 1/years. It can be concluded that the $\tau_0$ is approximately the same for the plots in Figs. 2-4.

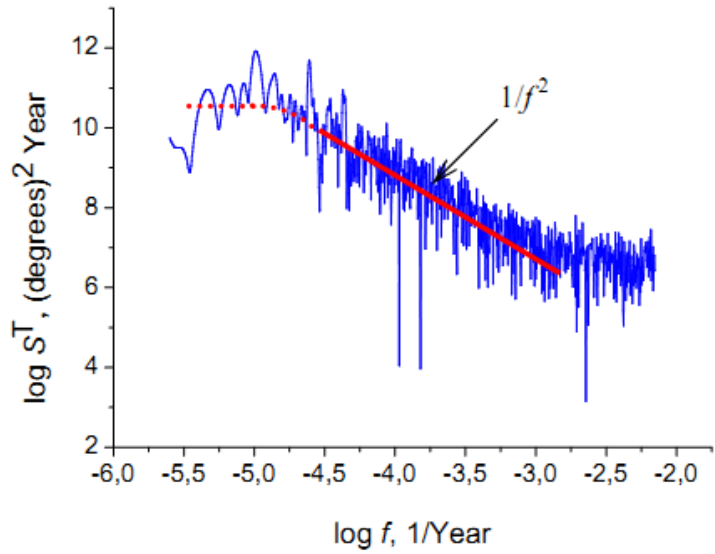

**Figure 5**. The frequency dependence of spectral density fluctuations, $S^T$ for temperature [9]. The solid and dotted lines are guidelines for the eye.

Figure 6 presents the spectral density fluctuations of the solar insolation [10].

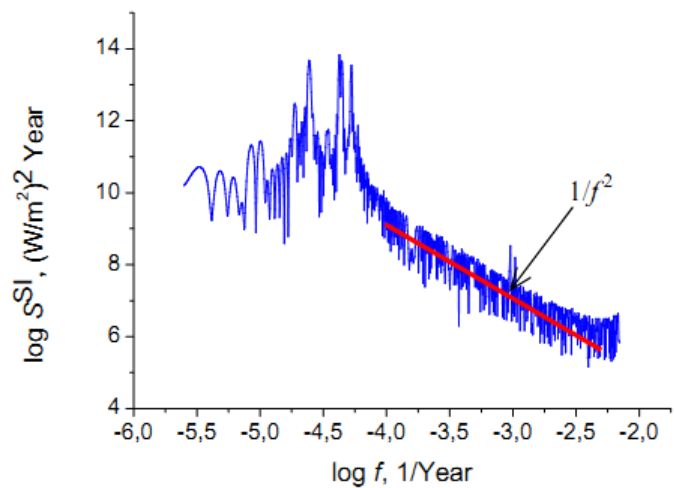

**Figure 6.** The frequency dependence of spectral density fluctuations, $S^{SI}$ for solar insolation [10]. The solid line is a guideline to the eye

The dependence $S^{SI} \sim 1/f^2$ is clearly traced up to frequencies of $\sim 10^{-4}$ 1/year. At lower frequencies, in the range $10^{-4} \leq f \leq 10^{-5}$ 1/year, sharp fluctuations of $S^{SI}$ are observed with several characteristic peaks. The presence of these peaks gives no way of determining the value of $\tau$ for the Lorentzian responsible for the observed $S^{SI} \sim 1/f^2$ dependence.

The above-mentioned random process with the same time constant $\tau_0 \sim \frac{1}{2\pi}(10^{4.5} - 10^{4.7})$ years for several paleoenvironmental records is certainly not universal. Figure 7 shows the frequency dependence of the spectral density fluctuations of the atmospheric radiocarbon $\delta^{14}C = \frac{^{14}C}{^{12}C}$ [11].

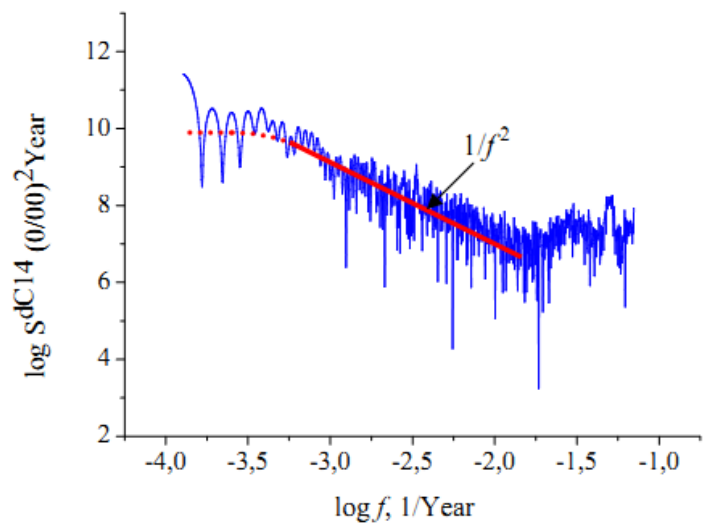

**Figure 7.** The frequency dependence of spectral density fluctuations for the atmospheric radiocarbon $\delta^{14}C$ [11].

As seen from Fig. 5, the characteristic value of $\tau_0$ for the process is $\sim \frac{1}{2\pi} 10^{3.3} \approx 300$ years.

Obviously, the most intriguing problem is the physical interpretation of the data, i.e., the identification of the random processes with a single time constant which are responsible for the appearance of single Lorentzians in the $S(f)$ dependences analyzed in this paper.

## 4. Conclusion

In this work, the time dependences of several important paleoclimatic parameters are considered for the first time as random processes ("noise"). Under this approach, it is shown that *single-time-constant random processes* (noise with a Lorentzian noise spectrum) play a very important and, perhaps, a decisive role in some important paleoclimatic dependences, such as the time dependences of the content of oxygen isotope ($\delta^{18}$O), atmospheric radiocarbon ($\Delta^{14}$C), deuterium ($^2$H), solar insolation, temperature, etc. For several of these processes, the characteristic time constants $\tau_0$ are very similar and equal to $(5\text{-}8) \times 10^3$ years. However, this value of $\tau_0$ is not universal. For example, the spectral density fluctuations of the atmospheric radiocarbon $\delta^{14}$C are characterized by a Lorentzian with $\tau_0 \approx 300$ years. At the same time, the frequency dependence of spectral density fluctuations $S^{\delta 18O}$ for benthic $\delta^{18}$O records contains two Lorentzians with $\tau_0 \approx 8000$ years and $\tau_0 > 10^5$ years.

To the best of our knowledge, this is the first indication of the role played by random processes in climate variations. At present, nothing is known about the physical nature of these random processes. Identification of the random processes responsible for the appearance of single Lorentzians in the *S(f)* dependences that were revealed in this work will allow a better understanding of the nature of climatic changes and, perhaps, of the near future of Humanity.